\documentclass[letter,longauth,traditabstract]{aa} 
\usepackage[]{graphicx}
\usepackage{natbib}
\bibpunct{(}{)}{;}{a}{}{,} 

\usepackage{txfonts}
\begin{document}
   \title{The $\beta$~Pictoris disk imaged by \emph{Herschel} PACS and SPIRE
     \thanks{\emph{Herschel} is an ESA space observatory with science
     instruments provided by European-led Principal Investigator
     consortia and with important participation from NASA.} 
   }

\author{
B.\,Vandenbussche\inst{1},
B.\,Sibthorpe\inst{2},
B.\,Acke\inst{1},
E.\,Pantin\inst{3},
G.\,Olofsson\inst{4},
C.\,Waelkens\inst{1},
C.\,Dominik\inst{5,6},
M.\,J.\,Barlow\inst{7},
J.\,A.\,D.\,L.\,Blommaert\inst{1},
J.\,Bouwman\inst{8},
A.\,Brandeker\inst{4},
M.\,Cohen\inst{9},
W.\,De\,Meester\inst{1},
W.\,R.\,F.\,Dent\inst{10},
K.\,Exter\inst{1}
J.\,Di\,Francesco\inst{11},
M.\,Fridlund\inst{12},
W.\,K.\,Gear\inst{13},
A.\,M.\,Glauser\inst{14,2},
H.\,L.\,Gomez\inst{13},
J.\,S.\,Greaves\inst{15},
P.\,C.\,Hargrave\inst{13},
P.\,M.\,Harvey\inst{16,17},
Th.\,Henning\inst{8},
A.\,M.\,Heras\inst{12},
M.\,R.\,Hogerheijde\inst{18},
W.\,S.\,Holland\inst{2},
R.\,Huygen\inst{1},
R.\,J.\,Ivison\inst{2,19},
C.\,Jean\inst{1},
S.\,J.\,Leeks\inst{20},
T.\,L.\,Lim\inst{20},
R.\,Liseau\inst{21},
B.\,C.\,Matthews\inst{11},
D.\,A.\,Naylor\inst{23},
G.\,L.\,Pilbratt\inst{12},
E.\,T.\,Polehampton\inst{20,22},
S.\,Regibo\inst{1},
P.\,Royer\inst{1},
A.\,Sicilia-Aguilar\inst{8},
B.\,M.\,Swinyard\inst{20},
H.\,J.\,Walker\inst{20},
R.\,Wesson\inst{7}
}

\institute{Instituut\ voor\ Sterrenkunde,\ Katholieke\ Universiteit\ Leuven,\ Celestijnenlaan\ 200\ D,\ B-3001\ Leuven,\ Belgium
\email{bart.vandenbussche@ster.kuleuven.be}
\and 
UK Astronomy Technology Centre, Royal Observatory Edinburgh, Blackford Hill, EH9 3HJ, UK\\
\and 
Laboratoire\ AIM,\ CEA/DSM-CNRS-Universit\'e\ Paris\ Diderot,\ IRFU/Service\ d’Astrophysique,\ Bˆat.709,\ CEA-Saclay,\ 91191\ Gifsur-Yvette\ C´edex,\ France
\and 
Department\ of\ Astronomy,\ Stockholm\ University,\ AlbaNova\ University\ Center,\ Roslagstullsbacken\ 21,\ 10691\ Stockholm,\ Sweden
\and 
Astronomical\ Institute\ Anton\ Pannekoek,\ University\ of\ Amsterdam,\ Kruislaan\ 403,\ 1098\ SJ\ Amsterdam,\ The\ Netherlands
\and 
Afdeling\ Sterrenkunde,\ Radboud\ Universiteit\ Nijmegen,\ Postbus\ 9010,\ 6500\ GL\ Nijmegen,\ The\ Netherlands
\and 
Department\ of\ Physics\ and\ Astronomy,\ University\ College\ London,\ Gower\ St,\ London\ WC1E\ 6BT,\ UK
\and 
Max-Planck-Institut\ f\"ur Astronomie,\ K\"onigstuhl\ 17,\ D-69117\ Heidelberg,\ Germany
\and 
Radio\ Astronomy\ Laboratory,\ University\ of\ California\ at\ Berkeley,\ CA\ 94720,\ USA
\and 
ALMA\ JAO,\ Av.\ El\ Golf\ 40\ -\ Piso\ 18,\ Las\ Condes,\ Santiago,\ Chile
\and 
National\ Research\ Council\ of\ Canada,\ Herzberg\ Institute\ of\ Astrophysics,\ 5071\ West\ Saanich\ Road,\ Victoria,\ BC,\ V9E\ 2E7,\ Canada
\and 
ESA Research\ and\ Science\ Support\ Department,\ ESTEC/SRE-S,\ Keplerlaan\ 1,\ NL-2201AZ,\ Noordwijk,\ The\ Netherlands
\and 
School\ of\ Physics\ and\ Astronomy,\ Cardiff\ University,\ Queens\ Buildings\ The\ Parade,\ Cardiff\ CF24\ 3AA,\ UK
\and 
Institute\ of\ Astronomy,\ ETH\ Zurich,\ 8093\ Zurich,\ Switzerland
\and 
School\ of\ Physics\ and\ Astronomy,\ University\ of\ St\ Andrews,\ North\ Haugh,\ St\ Andrews,\ Fife\ KY16\ 9SS,\ UK
\and 
Department\ of\ Astronomy,\ University\ of\ Texas,\ 1\ University\ Station\ C1400,\ Austin,\ TX 78712,\ USA
\and 
CASA,\,University\,of\,Colorado,\,389-UCB,\,Boulder,\,CO 80309,\,USA
\and 
Leiden\ Observatory,\ Leiden\ University,\ PO\ Box\ 9513,\ 2300\ RA,\ Leiden,\ The Netherlands
\and 
Institute\ for\ Astronomy,\ University\ of\ Edinburgh,\ Blackford\ Hill,\ Edinburgh\ EH9\ 3HJ,\ UK
\and 
Space\ Science\ and\ Technology\ Department,\ Rutherford\ Appleton\ Laboratory,\ Oxfordshire,\ OX11\ 0QX,\ UK
\and 
Department\ of\ Radio\ and\ Space\ Science,\ Chalmers\ University\ of\ Technology,\ Onsala\ Space\ Observatory,\ 439\ 92\ Onsala,\ Sweden
\and 
Institute for Space Imaging Science,\ University\ of\ Lethbridge,\ Lethbridge,\ Alberta,\ T1J\ 1B1,\ Canada
}

   \date{Received 31-March-2010; accepted 18-May-2010}

\abstract
  {We obtained \emph{Herschel} PACS and SPIRE images of the thermal emission
   of the debris disk around the A5V star $\beta$\,Pic.  The disk
   is well resolved in the PACS filters at 70, 100, and 160\,$\mu$m. 
   The surface brightness profiles
   between 70 and 160\,$\mu$m show no significant asymmetries along
   the disk, and are compatible with 
   90\% of the emission between 70 and 160\,$\mu$m originating in a region closer than 200\,AU
   to the star.  Although only marginally resolving the
   debris disk, the maps obtained in the SPIRE 250 -- 500\,$\mu$m filters 
   provide full-disk photometry, completing the SED
   over a few octaves in wavelength that had been previously inaccessible. The small
   far-infrared spectral index ($\beta = 0.34$) indicates that the grain size 
   distribution in the inner disk ($<$200\,AU) is inconsistent with a local
   collisional equilibrium. The size distribution is either
   modified by non-equilibrium effects, or exhibits a wavy pattern, caused by an 
   under-abundance of 
   impactors which have been removed by radiation pressure.
  }

\authorrunning{B. Vandenbussche et al.}
\titlerunning{The $\beta$\,Pic debris disk imaged by \emph{Herschel}}

\maketitle

%

\section{Introduction}

The $\beta$\,Pic disk, discovered by IRAS \citep{aumann:1984},
was the first debris disk to be directly imaged in scattered 
light \citep{smith:1984}.  It is seen close to edge-on and extends in the optical 
out to 95\arcsec, corresponding to 1800\,AU \citep{larwood:2001}.

$\beta$\,Pic (A5V) is one of the closest \citep[19.44\,$\pm$\,0.05\,pc,][]{vanleeuwen:2007}
and youngest debris disks.  The estimated age \citep[12 Myr,][]{zuckerman:2001}  
significantly  exceeds typical timescales for the survival of pristine
circumstellar dust grains \citep[e.g.,][]{fedele:2010}, hence 
continuous replenishment of the dust, presumably through collisions of planetesimals, is needed.  
The closeness of the object ensures that it can also be spatially resolved at long wavelengths: 
\citet{holland:1998} resolved the disk at 850\,$\mu$m and \citet{liseau:2003} at 1200\,$\mu$m.

Optical and near-infrared observations of the inner part ($<$100AU) of the disk yield
evidence of asymmetries such as warps and density contrasts, which may relate to
the presence of planetesimals \citep{kalas:1995,pantin:1997,mouillet:1997,heap:2000, telesco:2005}.
\citet{lagrange:2009B} imaged a possible companion at a projected distance of 8\,AU from the star. 

Images of $\beta$\,Pic in scattered stellar light directly detect small 
grains and indirectly larger grains that produce the smaller ones through 
collisions.   The grain-size distribution can be quantitatively constrained from the 
spectral energy distribution (SED) of the disk, in the infrared and (sub)mm domains.  
The spectral index of the SED at the longest wavelengths \citep{liseau:2003,nilsson:2009} 
is inferred to be fairly low, which according to modeling 
\citep{draine:2006,natta:2007,ricci:2010} can be interpreted as a deficit of 
small grains.

In this paper, we present far-infrared imaging of the $\beta$\,Pic debris disk in six \emph{Herschel}
photometric bands between 70 and 500\,$\mu$m.  These bands cover for the first time the 
long-wavelength side of the peak in the thermal emission of the disk, and
the large aperture of the telescope enables us to resolve the disk at far-IR wavelengths 
for the first time.  With these data, we measure the surface brightness profiles of the disk and 
readdress the issue of the grain-size distribution in the inner 200\,AU.


\section{Observations and data reduction}

We obtained maps of $\beta$\,Pic with the PACS and SPIRE 
instruments of \emph{Herschel} \citep{pilbratt:2010}.
The in-orbit performance, scientific capabilities, 
calibration methods, and accuracy are outlined by
\citet{poglitsch:2010} for the PACS instrument and by
\citet{griffin:2010} and \citet{swinyard:2010} for the
SPIRE instrument.
The observations were carried out during the science demonstration phase 
as proposed in the `Stellar Disk Evolution' guaranteed time proposal 
(PI G. Olofsson).    
Table~\ref{table:observations} gives a summary of the observations.
The deep PACS observation at 70$\mu$m and 160$\mu$m is a standard PACS 
photometer scan map, split into a scan and cross-scan on the sky. The sky 
scan speed
was 10\arcsec\,sec$^{-1}$. The homogeneously covered area of the deep map is 
$2.5\arcmin \times 2.5\arcmin$.  
The observation at 100\,$\mu$m is much shallower, with a single
scan direction at a rate of 20\arcsec\,sec$^{-1}$, homogeneously covering an area of 
$2\arcmin \times 2\arcmin$.
The PACS beams at 70, 100, and 160\,$\mu$m are 
5.6, 6.8, and 11.3\arcsec\,FWHM.
In the SPIRE observation, the three bands are observed simultaneously 
in a standard scan map. The map coverage is $8\arcmin \times 8\arcmin$.
The SPIRE FWHM beam sizes in the 250, 350, and 500\,$\mu$m channels are
18.1, 25.2, and 36.9\arcsec\ respectively.

\onltab{1}{
\begin{table}
\caption{Observation log}
\label{table:observations}
\centering                         
\begin{tabular}{c c c c c c}        
\hline\hline                 
           & observation& date       & Pos Angle  & duration & filters  \\
\hline
SPIRE      & 1342187327 & 2009-11-30 & 154.96\degr & 3336 s & 250,350,500 \\
PACS       & 1342185457 & 2009-10-07 & 106.54\degr & 866  s &  100,160 \\
PACS       & 1342186613 & 2009-11-01 & 130.98\degr & 5506 s &  70, 160 \\
PACS       & 1342186612 & 2009-11-01 & 130.98\degr & 5506 s &  70, 160 \\
\hline     
\end{tabular}
\end{table}
}


The data processing is described in Appendix.
The absolute flux calibration accuracy of the resulting PACS maps is better 
than 10\% at 70 and 100\,$\mu$m, and 20\% at 160\,$\mu$m \citep{poglitsch:2010}.
The flux calibration accuracy of the SPIRE maps is better than 15\%.
\citep{swinyard:2010}. 
The 1$\sigma$ noise levels of the maps are listed in 
Table~\ref{table:measuredQuantities}.


\section{Analysis}

\begin{figure*}
 \centering
 \includegraphics[width=60mm]{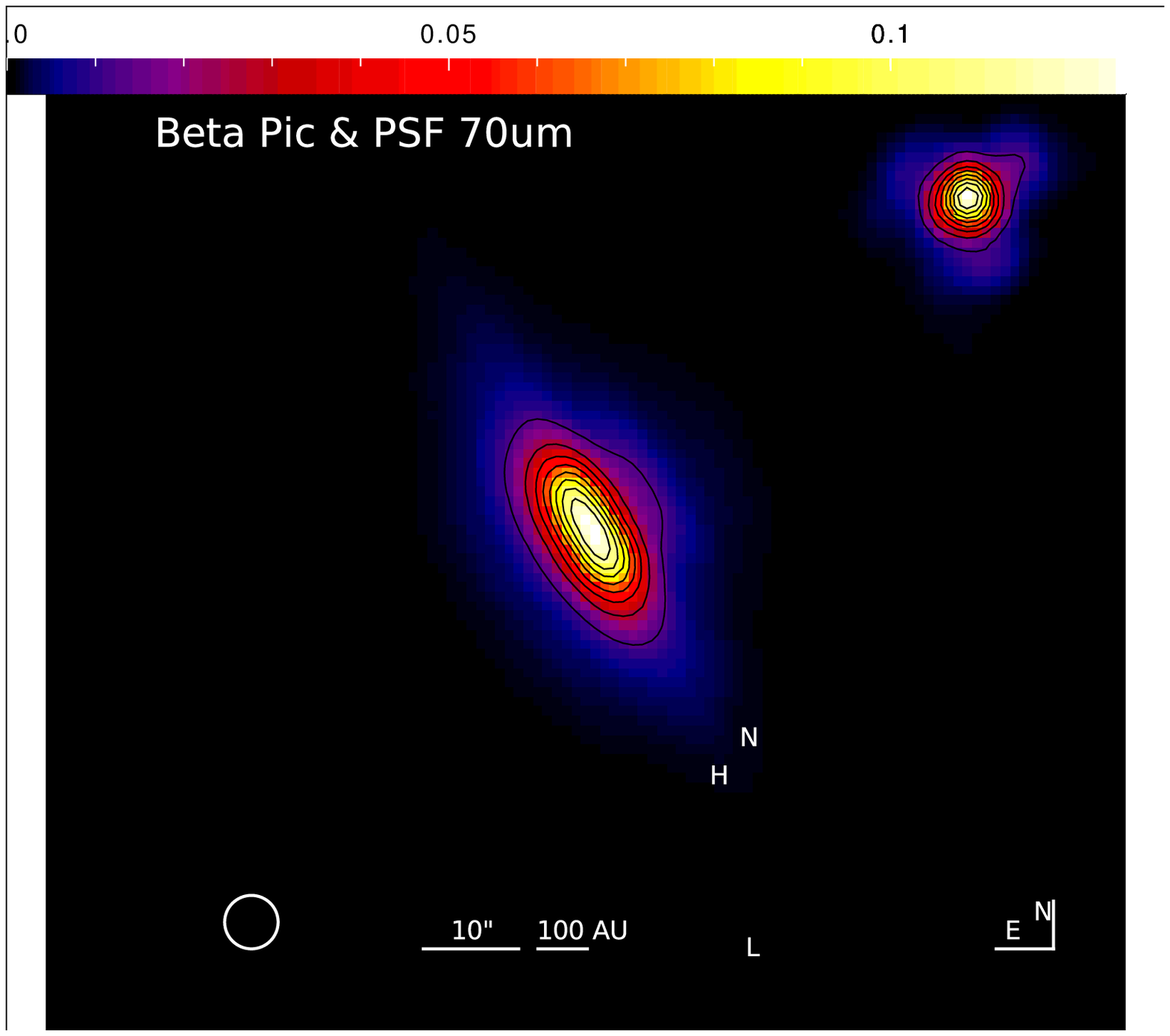}
 \includegraphics[width=60mm]{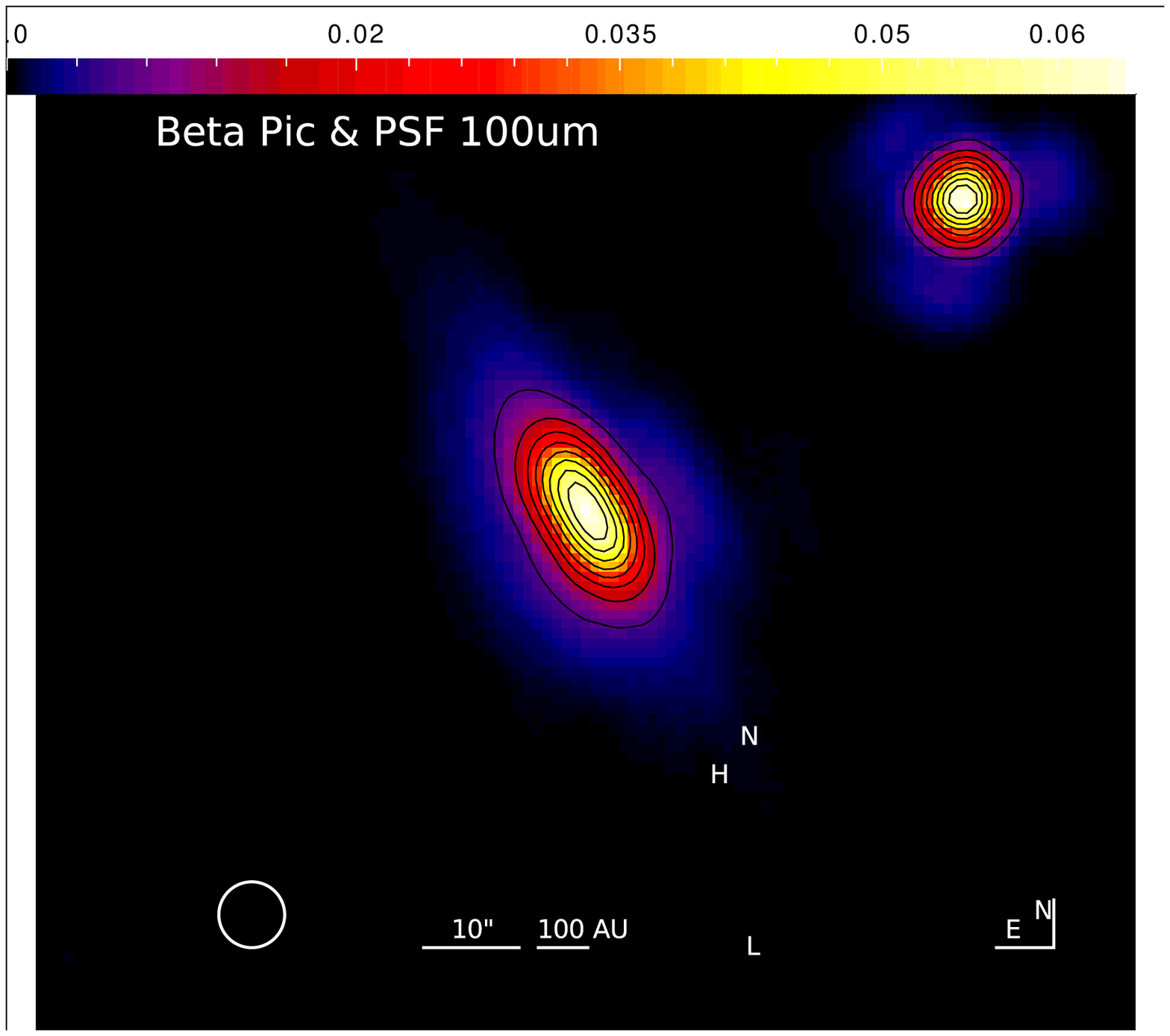}
 \includegraphics[width=60mm]{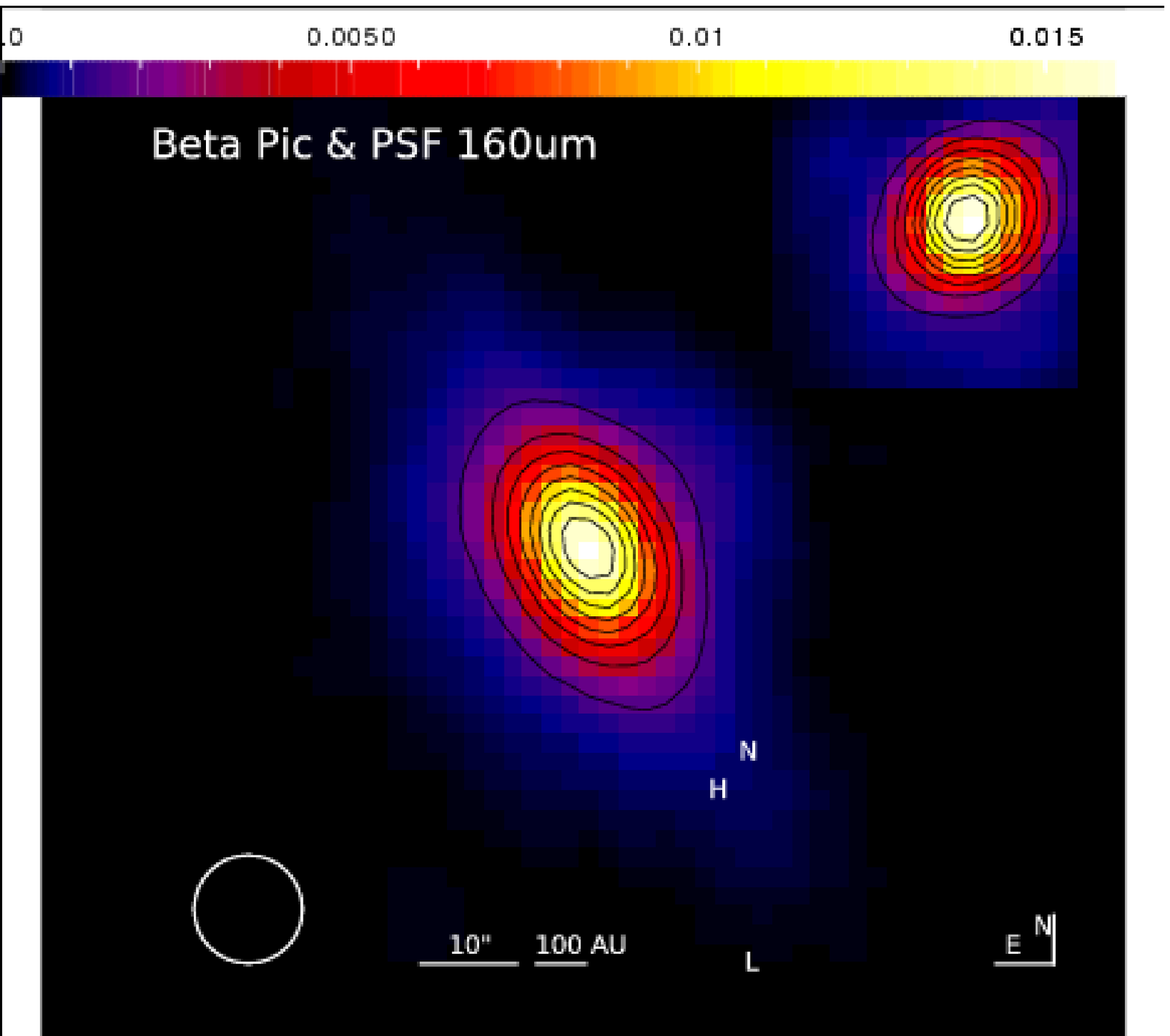}
 \includegraphics[width=60mm]{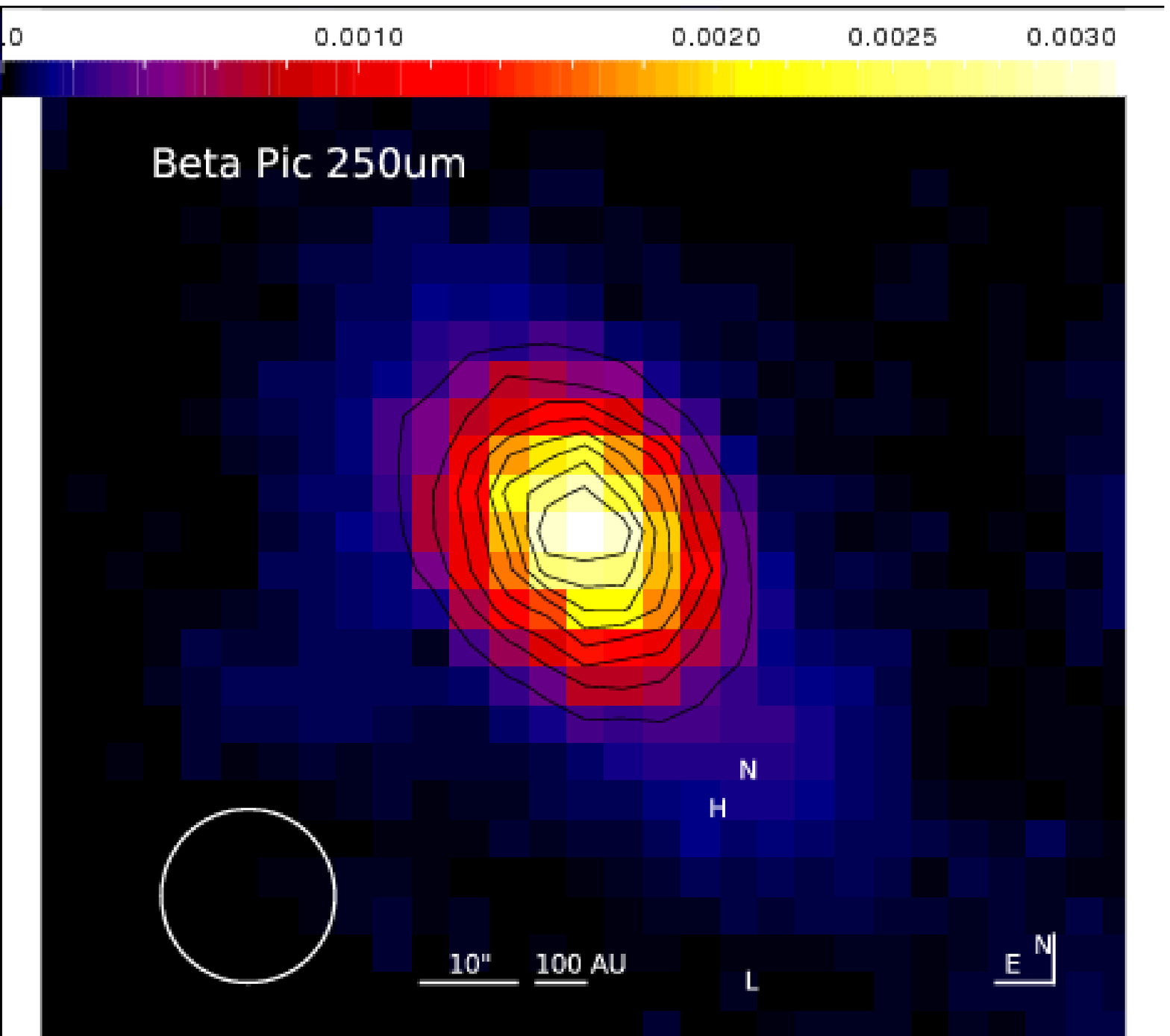}
 \includegraphics[width=60mm]{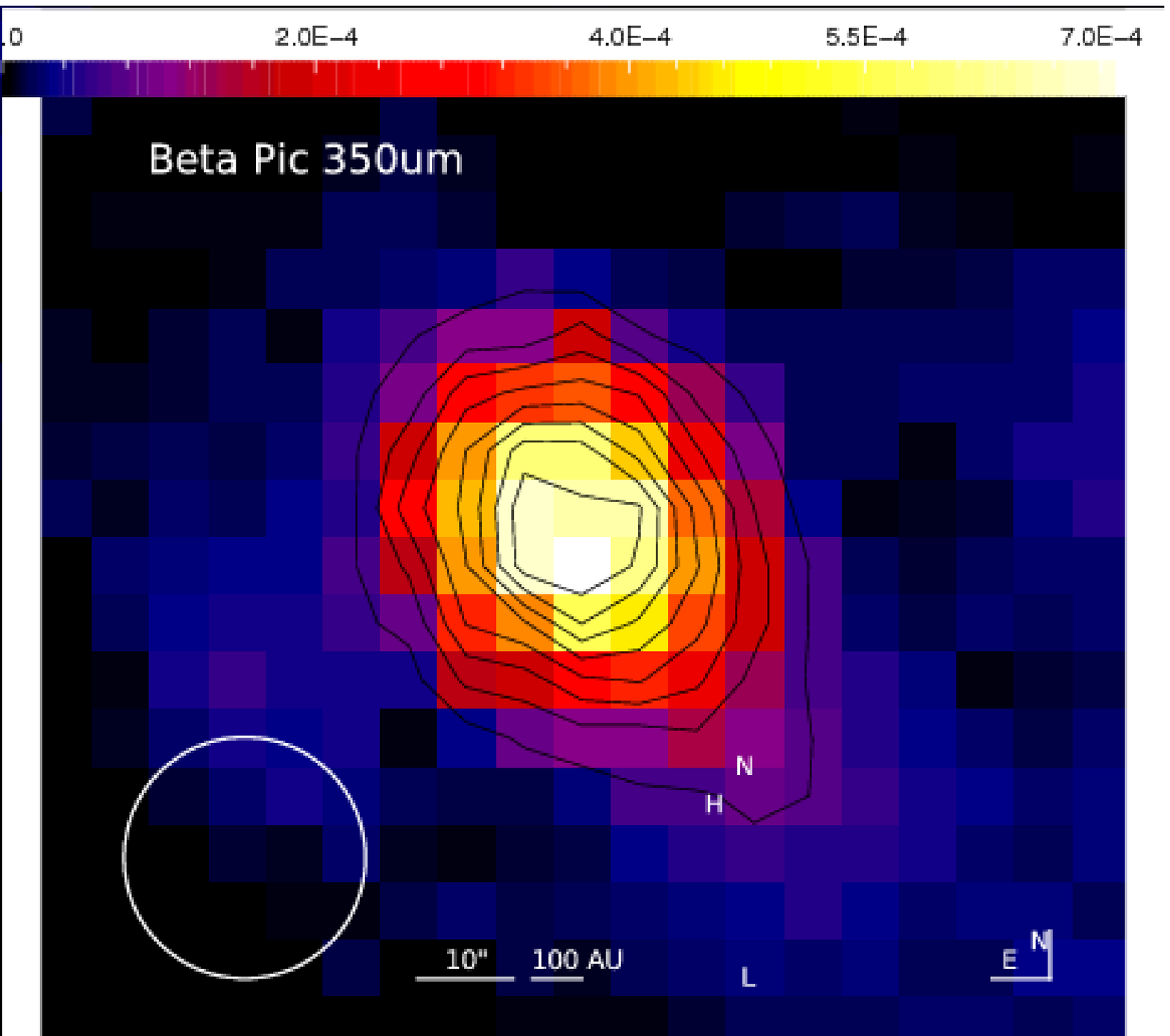}
 \includegraphics[width=60mm]{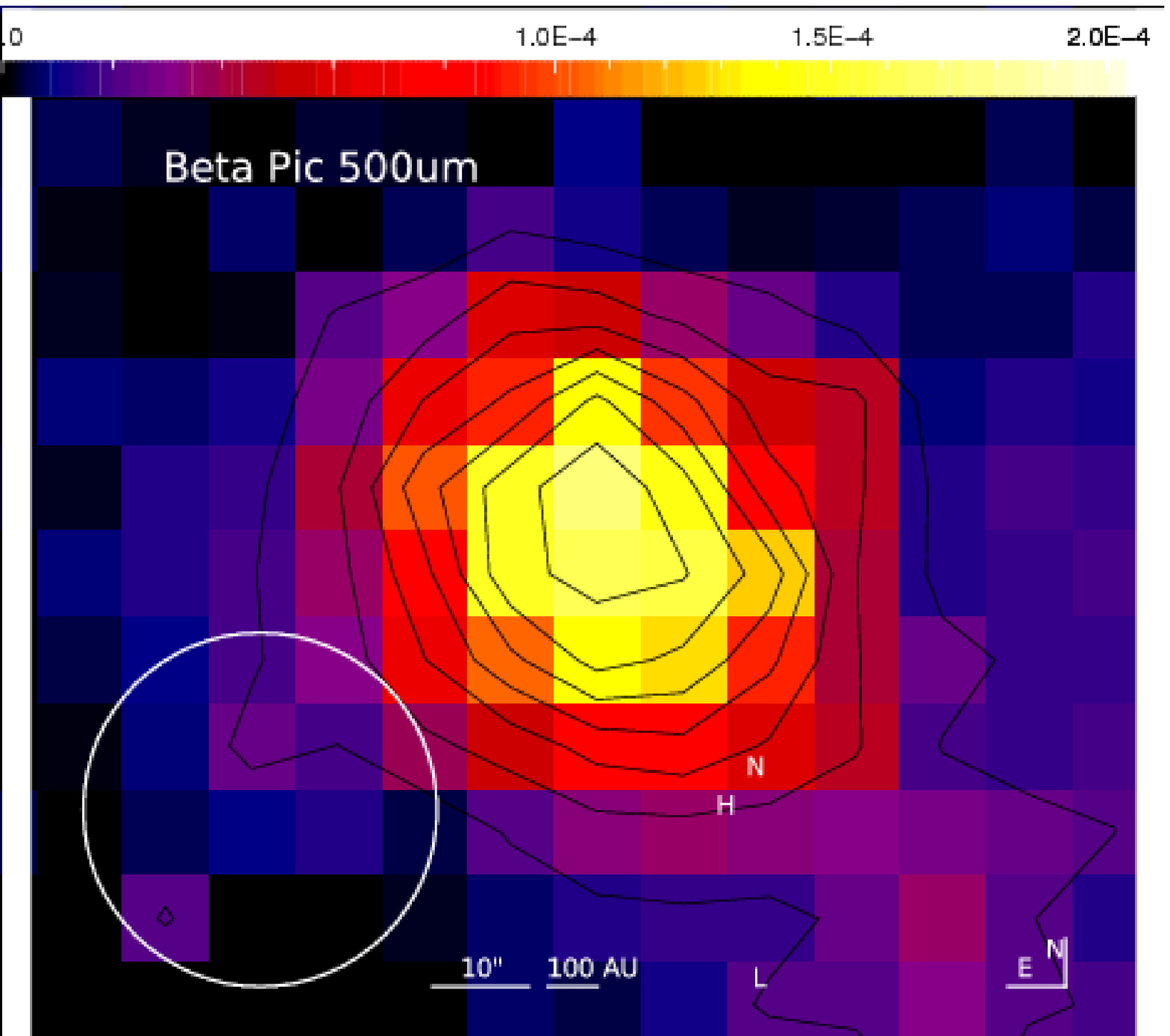}
 \caption{Surface brightness maps of the $\beta$\,Pic debris disk at 
 70, 100, 160, 250, 350, and 500\,$\mu$m.  The PACS PSFs, rotated 
 to match the position angle of the telescope
 at the time of the $\beta$\,Pic observations are depicted in the upper 
 right corner of the images.  
 The SPIRE PSFs are depicted in 
 Fig.~\ref{Fig:spirePsfs}.
 All images are scaled linearly, contour lines are in steps of 10$\%$ of 
 the peak flux.  The surface brightness unit is Jy\,arcsec$^{-2}$.
 The white circle shows the beam FWHM.  The position of
 the flux peaks observed at 850, 870, and 1200\,$\mu$m by
 \citet{holland:1998}, \citet{nilsson:2009}, and \citet{liseau:2003}
 are indicated with H, N, and L.}
 \label{Fig:betapicmaps}
\end{figure*}

In Fig.~\ref{Fig:betapicmaps}, we show the maps obtained in the three PACS 
filters (70, 100, and 160\,$\mu$m) and the three SPIRE filters 
(250, 350, and 500\,$\mu$m).  We also compare the point spread functions (PSFs) measured
on the asteroid Vesta using the same satellite scan speed, processed 
as the $\beta$\,Pic maps and rotated to align with the telescope pupil 
orientation on the sky during the $\beta$\,Pic observations as listed
in online Table~\ref{table:observations}.

These images show a clearly resolved disk from 70--160$\mu$m.
Each map was fitted using a 2D Gaussian function.  
Within the 2\arcsec\ \emph{Herschel} pointing accuracy, the Gaussian center
matches the star's optical position.
The fitted position angles, listed in Table~\ref{table:measuredQuantities}, 
agree with the optical disk position angle of 30\fdg8 reported by \citet{kalas:1995}.
Cross-sections orthogonal to the disk position angle in the NW to SE direction 
show no significant broadening compared to the PSF. The disk is not 
resolved in the vertical direction.
The feature towards the NW, visible in the 70--160\,$\mu$m
images, is produced by the three-lobed PACS PSF.

\begin{table}
\caption{Overview of measured quantities: position angle PA, northeast (NE) and 
southwest (SW) extent (signal reaching the 1\,$\sigma$ noise), 
map noise level, beam FWHM, and the flux density integrated over a 60\arcsec\ aperture.
}
\label{table:measuredQuantities}
\centering                         
\begin{tabular}{c c c c c c c}        
\hline\hline                 
$\lambda$   & PA      & NE        & SW       & 1$\sigma$ noise         & beam      & $F_\nu$ \\ 
($\mu$m)    & (\degr)  & (\arcsec) & (\arcsec)& (mJy\, \arcsec$^{-2}$)  & (\arcsec) & (Jy)    \\ 
\hline
70\,$\mu$m  & 29\fdg9 & 68        & 67       & 0.079     & 5.6       & 16.0\,$\pm$\,0.8 \\
100\,$\mu$m & 30\fdg3 & 55        & 56       & 0.086     & 8.6       &  9.8\,$\pm$\,0.5 \\
160\,$\mu$m & 28\fdg1 & 63        & 60       & 0.044     & 11.3      &  5.1\,$\pm$\,0.5 \\
250\,$\mu$m &         & 62        & 72       & 0.015     & 18.1      &  1.9\,$\pm$\,0.1 \\
350\,$\mu$m &         & 42        & 83       & 0.007     & 25.2      & 0.72\,$\pm$\,0.05 \\
500\,$\mu$m &         & 33        & 80       & 0.004     & 36.9      & 0.38\,$\pm$\,0.03 \\
\hline     
\end{tabular}
\end{table}

\onlfig{6}{
\begin{figure*}
 \centering
 \includegraphics[width=60mm]{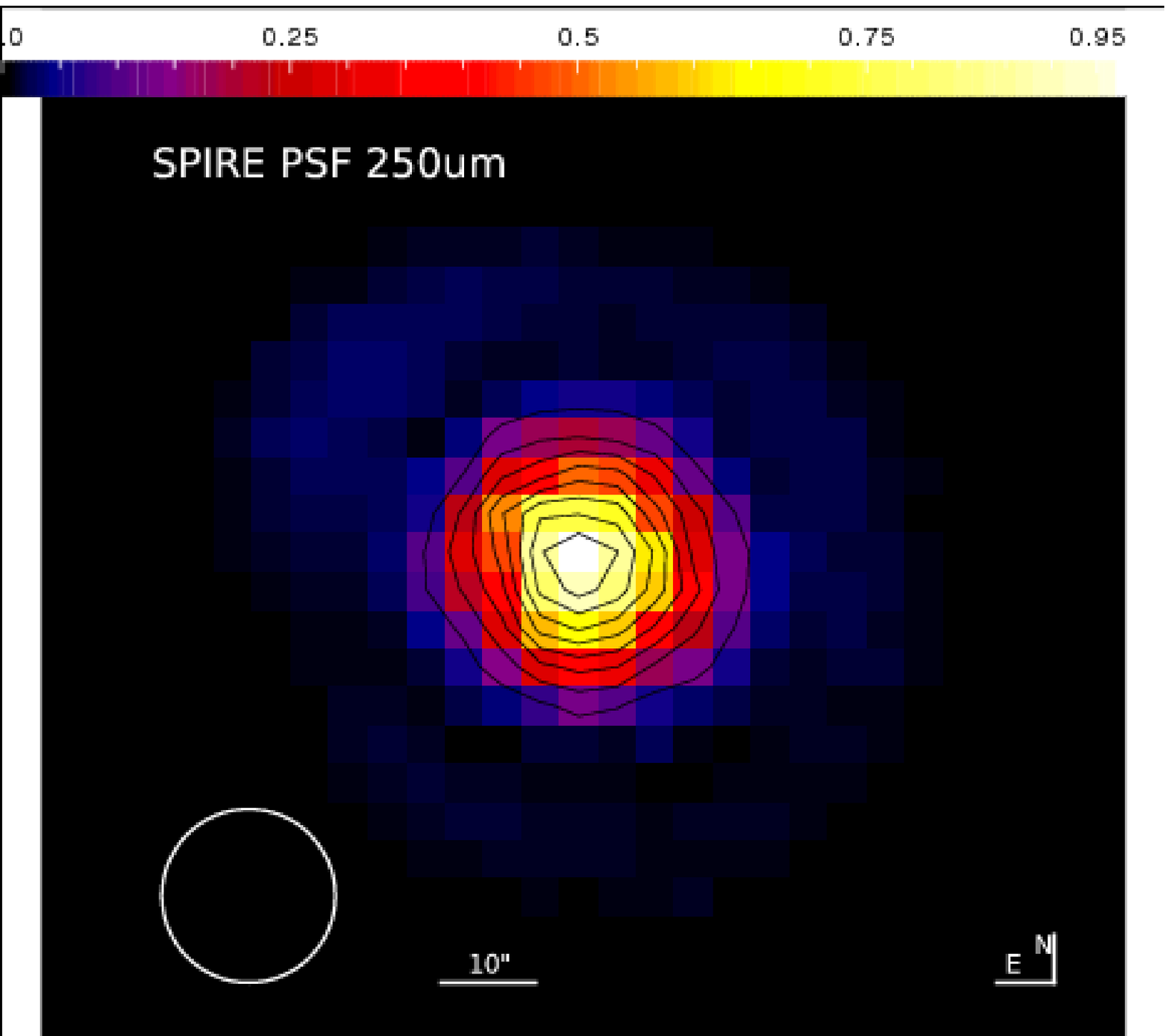}
 \includegraphics[width=60mm]{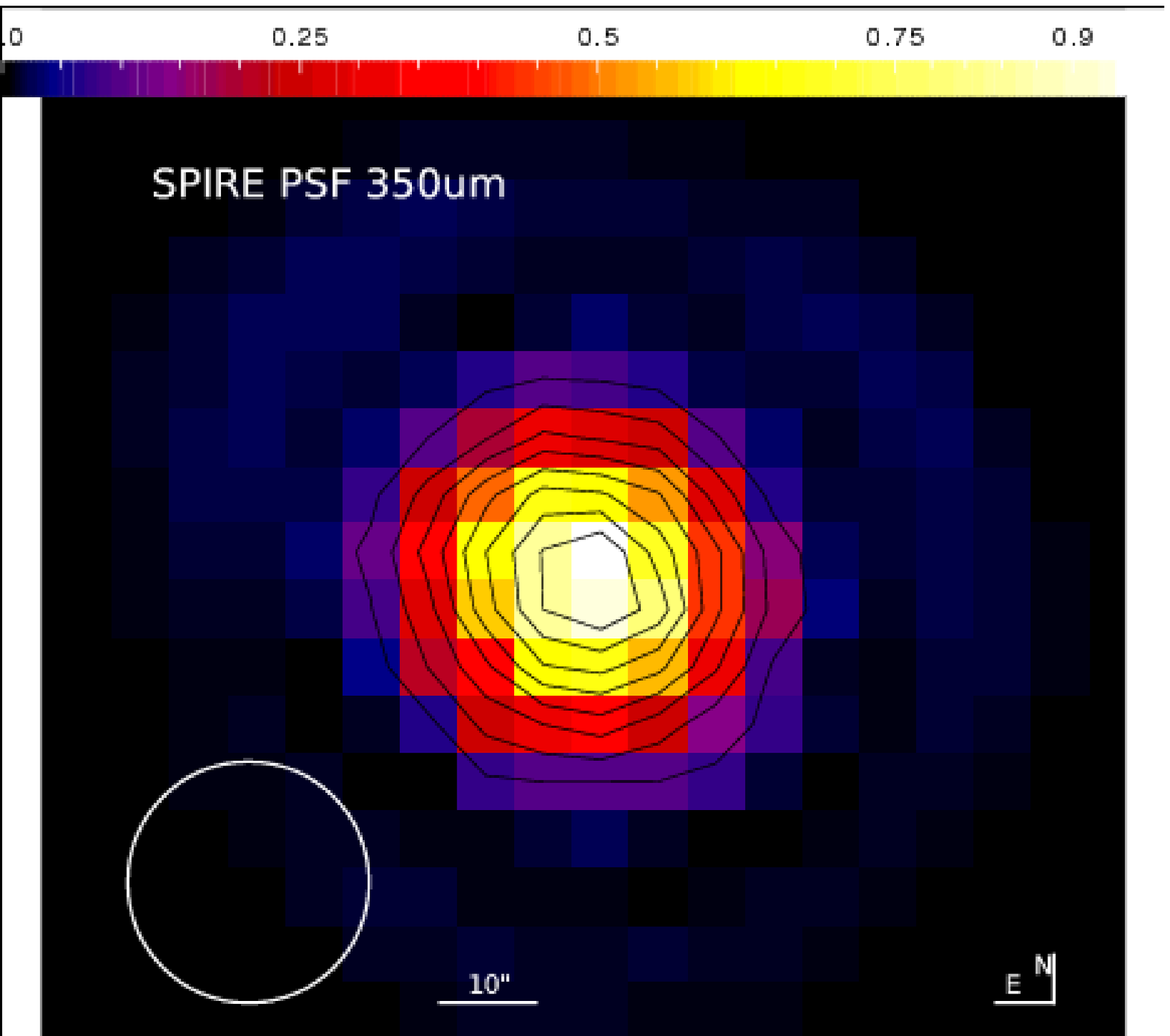}
 \includegraphics[width=60mm]{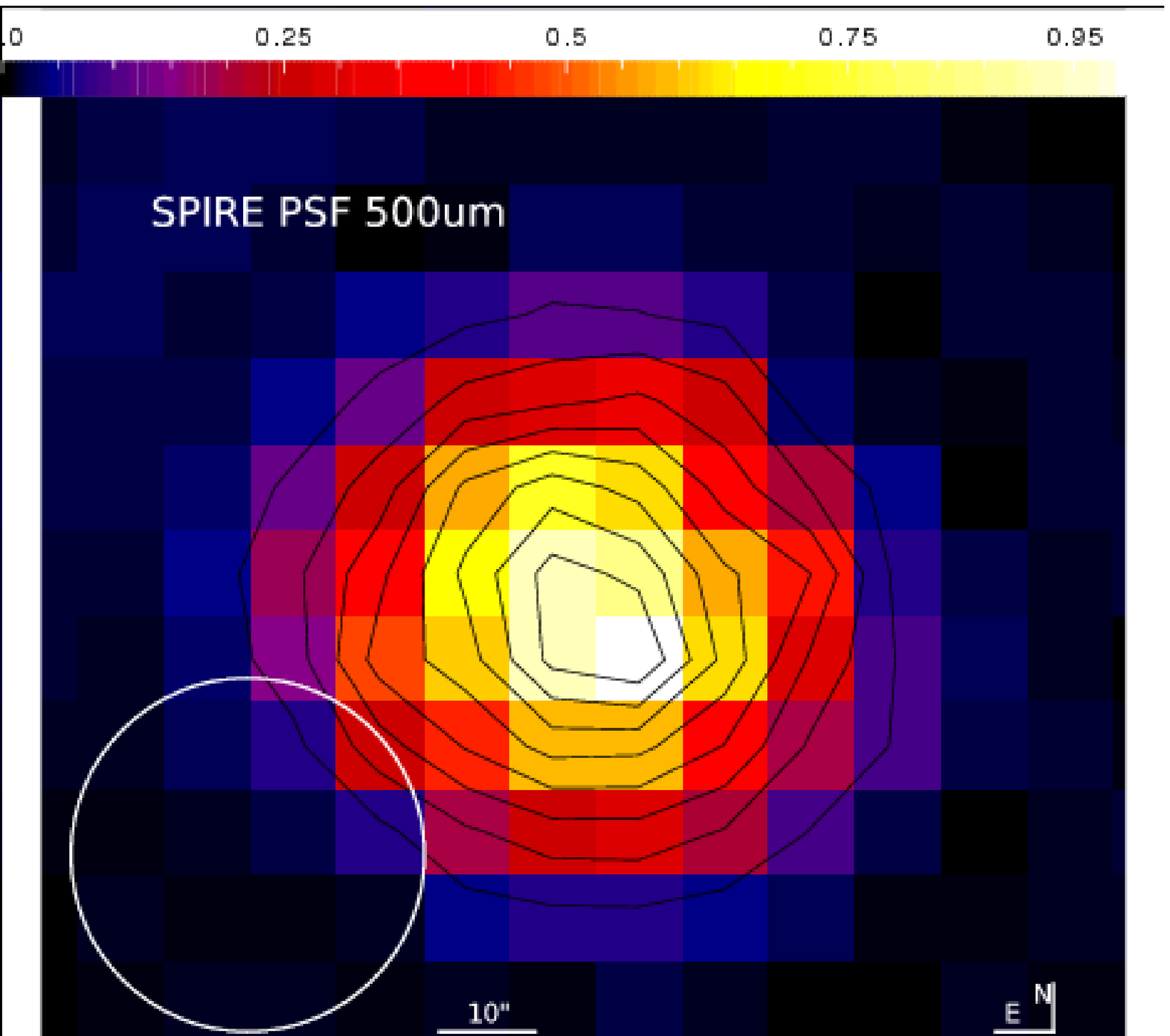}
 \caption{The 250, 350 and 500\,$\mu$m SPIRE PSFs, rotated to match
 the position angle at the time of the $\beta$\,Pic observations.
 The PSF images are scaled linearly, contour lines are in steps of 10$\%$ of
  the peak flux. The white circle shows the beam FWHM.}
 \label{Fig:spirePsfs}
\end{figure*}
}

\begin{figure*}
 \centering
 \includegraphics[width=51mm]{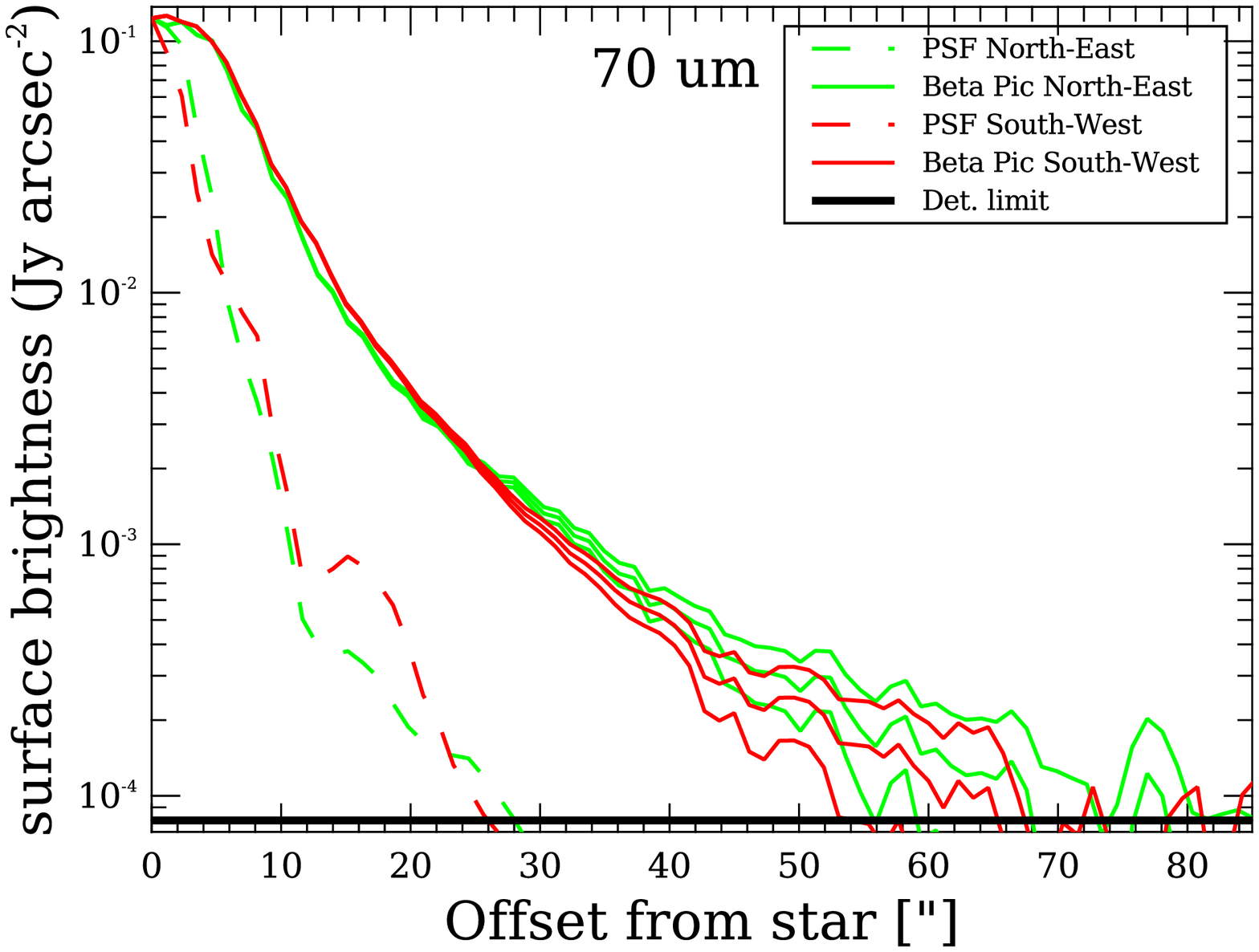}
 \includegraphics[width=51mm]{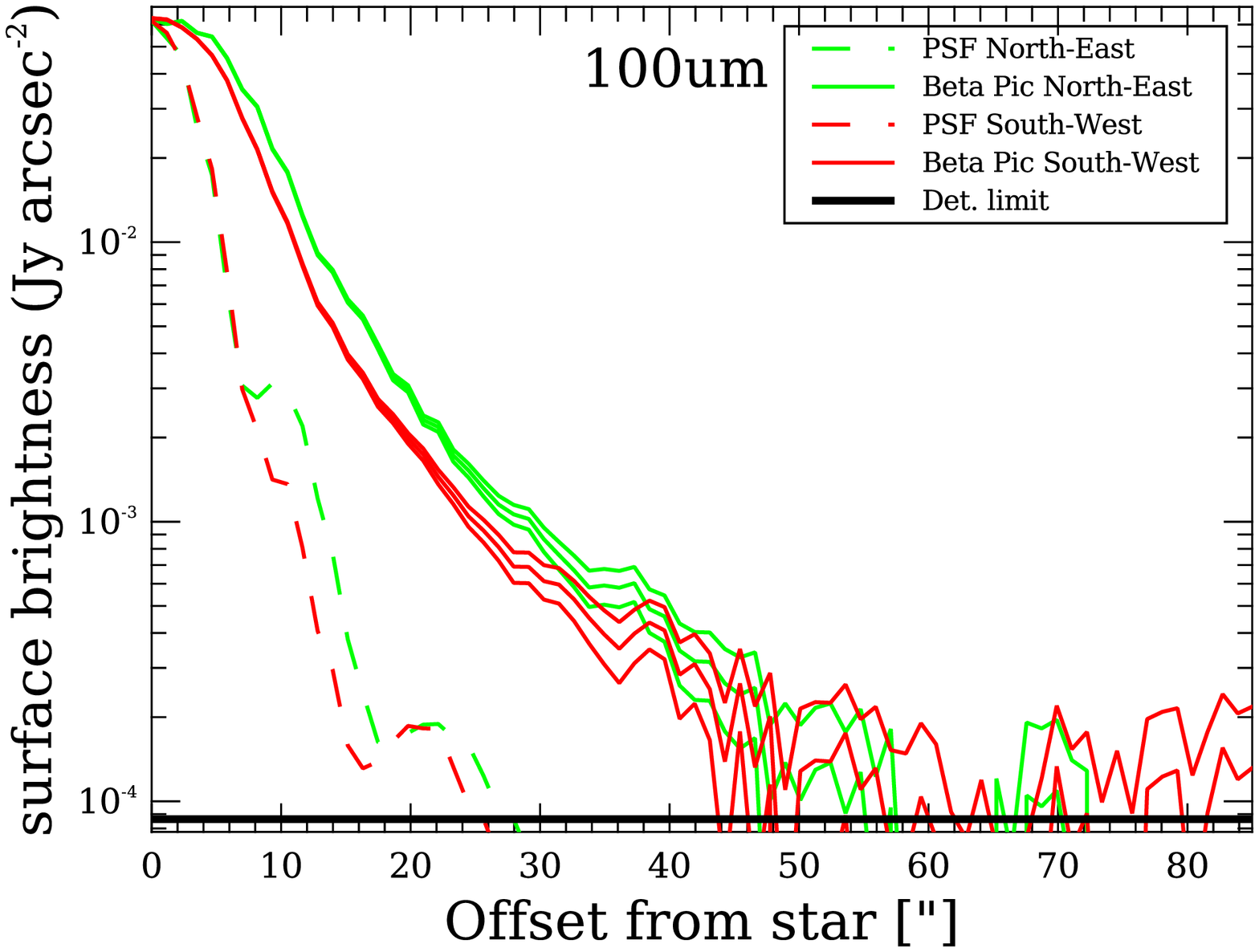}
 \includegraphics[width=51mm]{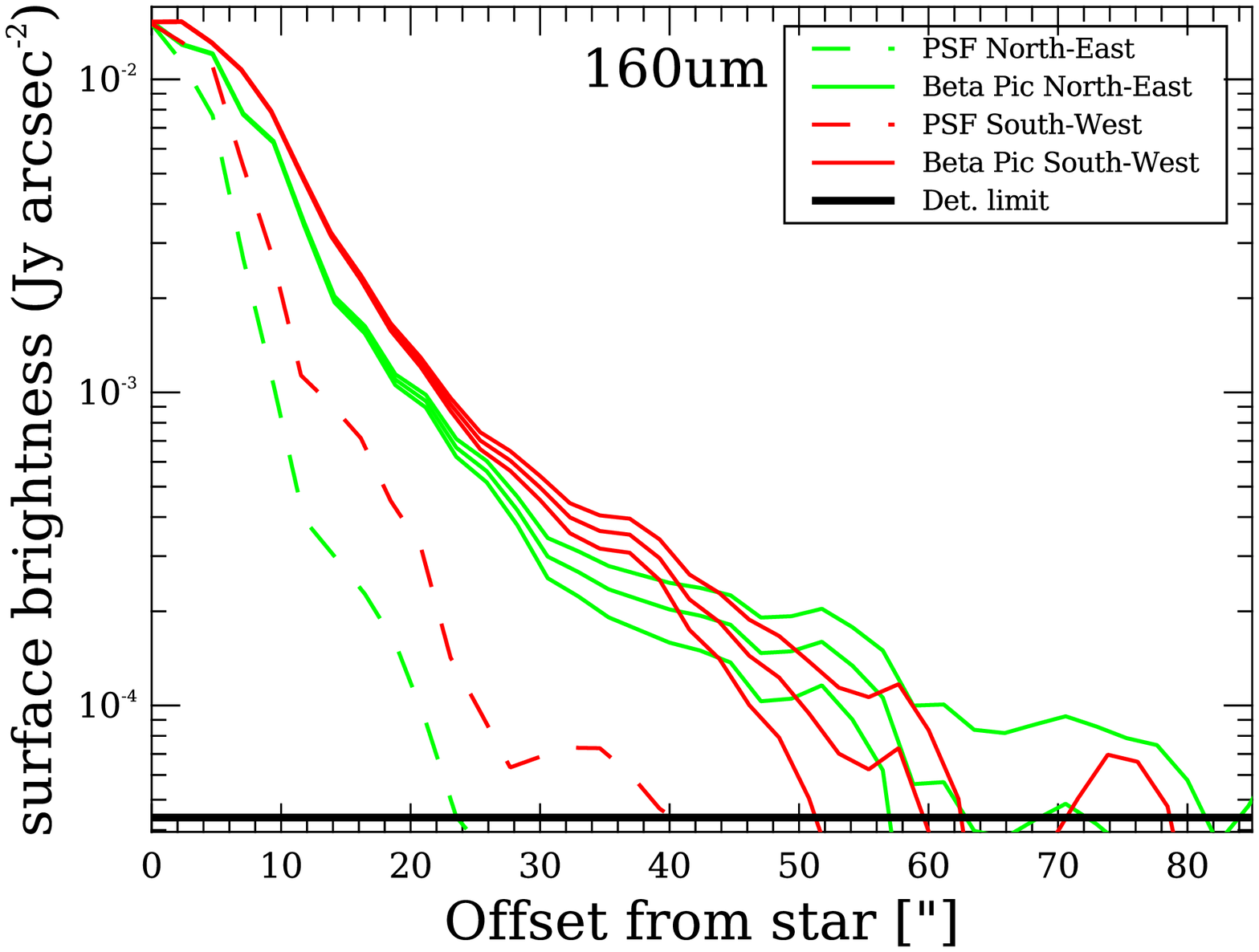}
 \includegraphics[width=51mm]{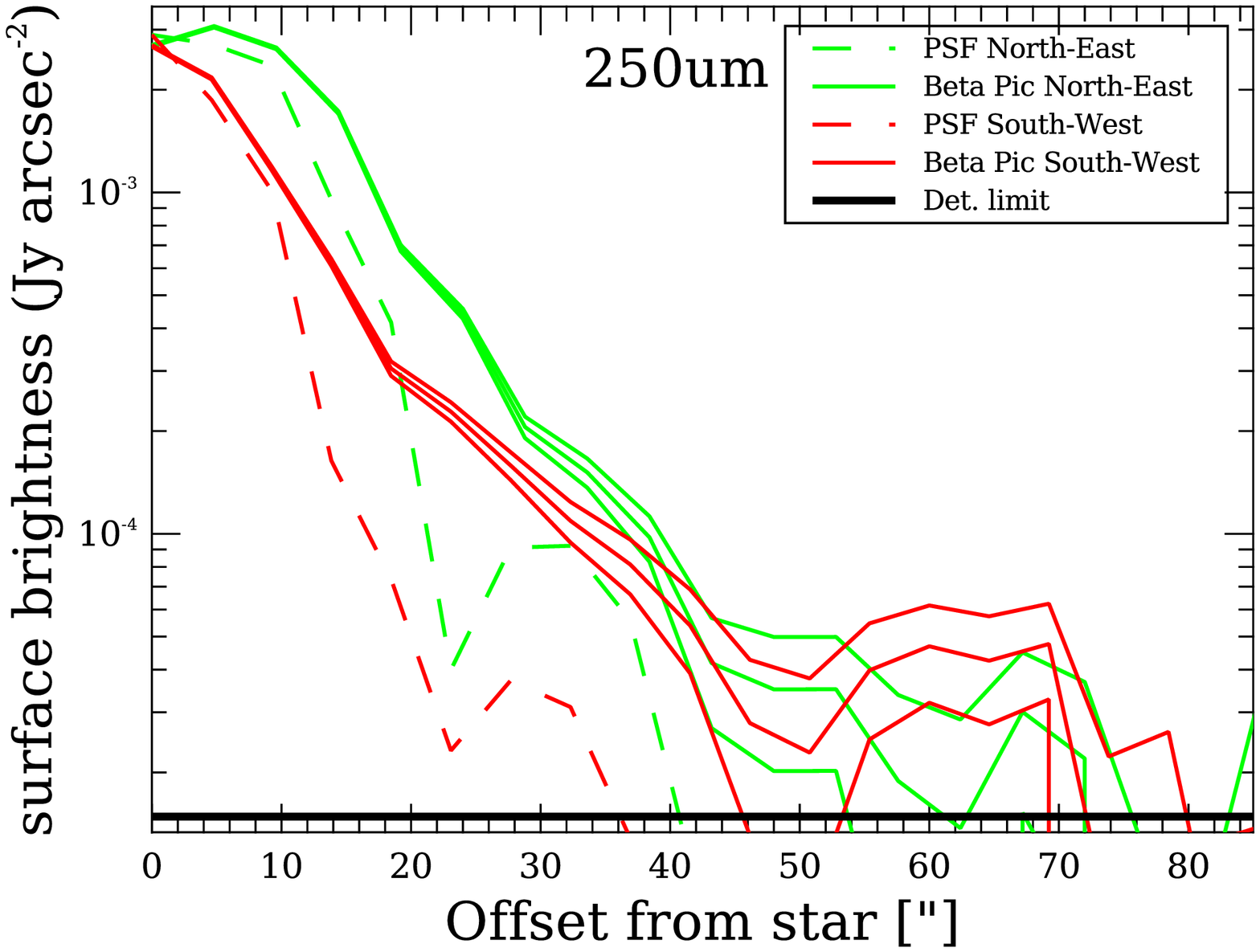}
 \includegraphics[width=51mm]{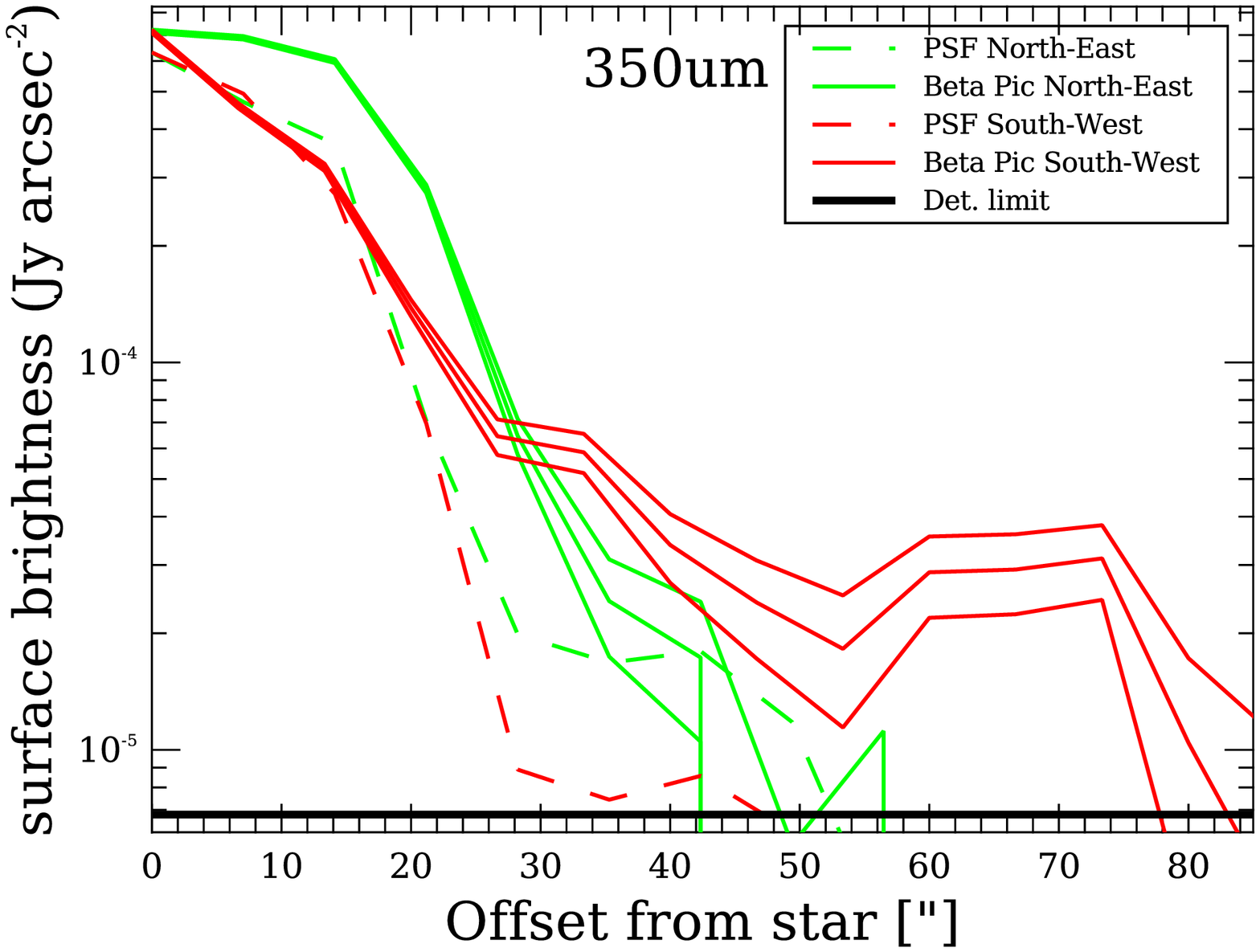}
 \includegraphics[width=51mm]{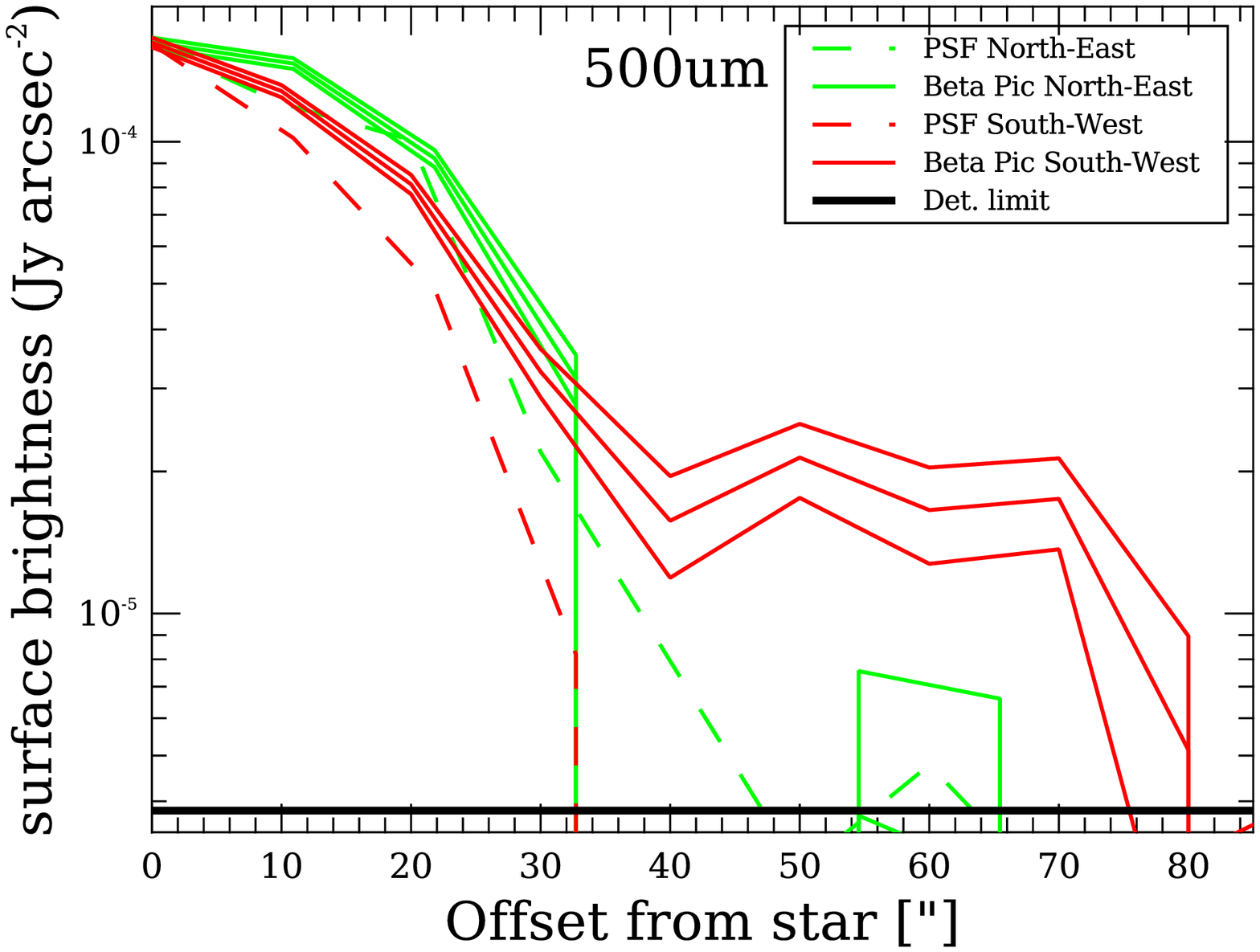}

 \caption{Surface brightness profiles along the disk in NE-SW 
          direction, following the 30.8\degr\ position angle. The black horizontal
	  line shows the 1$\sigma$ detection limit in the $\beta$\,Pic maps.
	  The $\beta$\,Pic profiles are shown with the 1$\sigma$ errors. 
	  }
 \label{Fig:profile_NESW}
\end{figure*}

In Fig.~\ref{Fig:profile_NESW}, we present the surface brightness profiles along
the disk position angle.  We compare them with the cross-sections aligned in the same
direction through the PSFs.  At 250 and 350\,$\mu$m, the disk is marginally 
resolved.  At 500\,$\mu$m, the
$\beta$~Pic profile shows no significant departure from the PSF profile,
with the exception of a cold blob in the southwest.
As can be seen in Fig.~\ref{Fig:betapicmaps},
the location of this feature in the 250 -- 500\,$\mu$m maps coincides with 
the flux peaks seen at 850 and 870$\mu$m by \citet{holland:1998} 
and \citet{nilsson:2009}, respectively.
However, the 100\,arcmin$^2$ region around $\beta$~Pic (depicted
in online Fig~\ref{Fig:backgroundsources}) shows
more than 50 background sources comparable to this feature in the
250\,$\mu$m map. The feature is therefore probably 
a background source.

\onlfig{2}{
\begin{figure*}
 \centering
 \includegraphics[width=16cm]{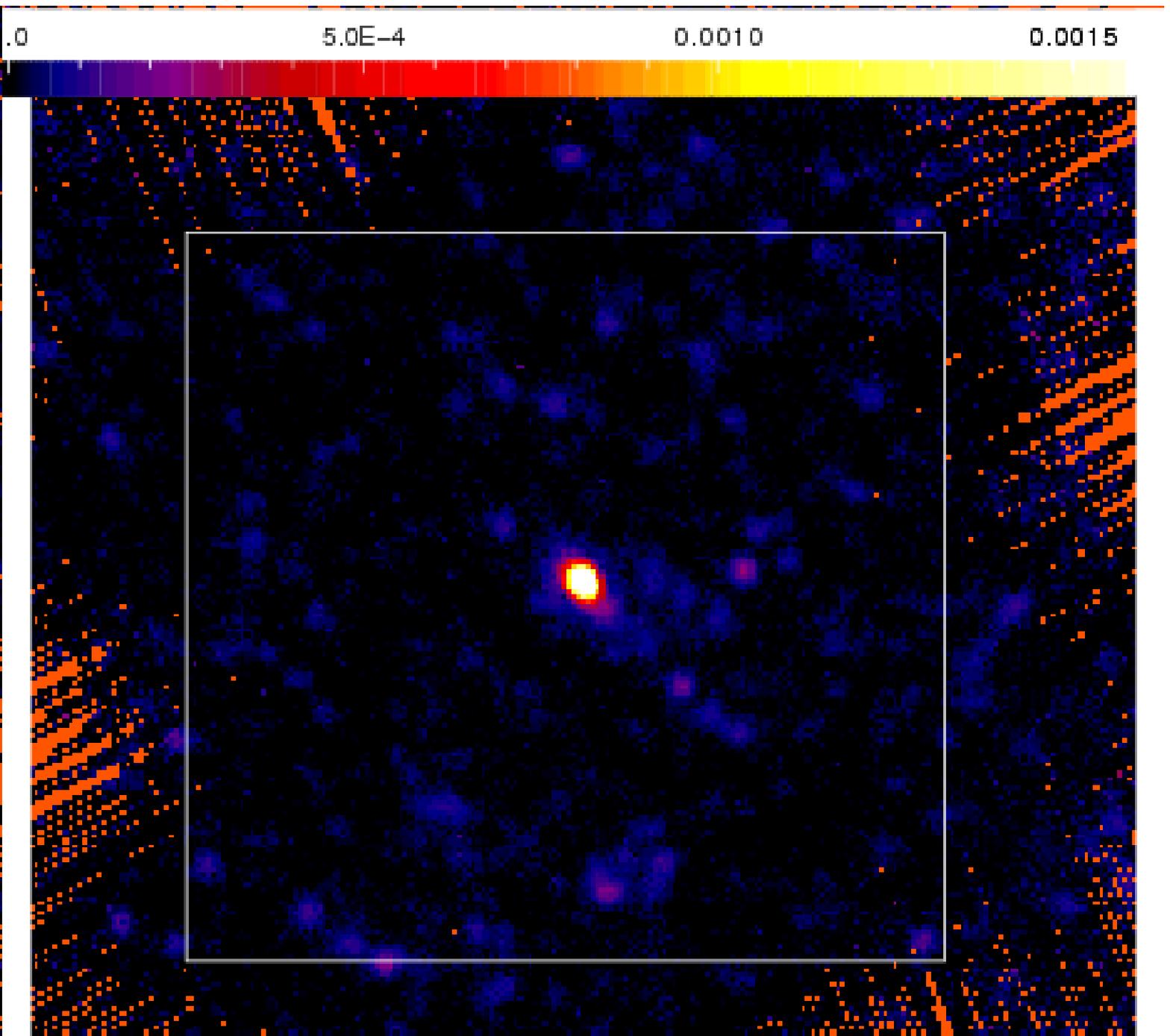}
 \caption{The 250\,$\mu$m SPIRE map around the $\beta$~Pic disk. The
 10x10\arcmin region delimited by the white square shows 
 more than 50 background sources comparable to the cold blob seen in the
 southwest of the disk.}
 \label{Fig:backgroundsources}
\end{figure*}
}

Other asymmetries between the northeast and southwest profile are within the
errors induced by the asymmetry of the PSF.
No sharp disk edge is seen; in all filters, the surface brightness
declines gradually to the 1$\sigma$ detection limit of the maps.
Table~\ref{table:measuredQuantities} lists the extent of the
detected emission region in the NE-SW direction.

\begin{figure}
 \centering
 \includegraphics[width=8cm]{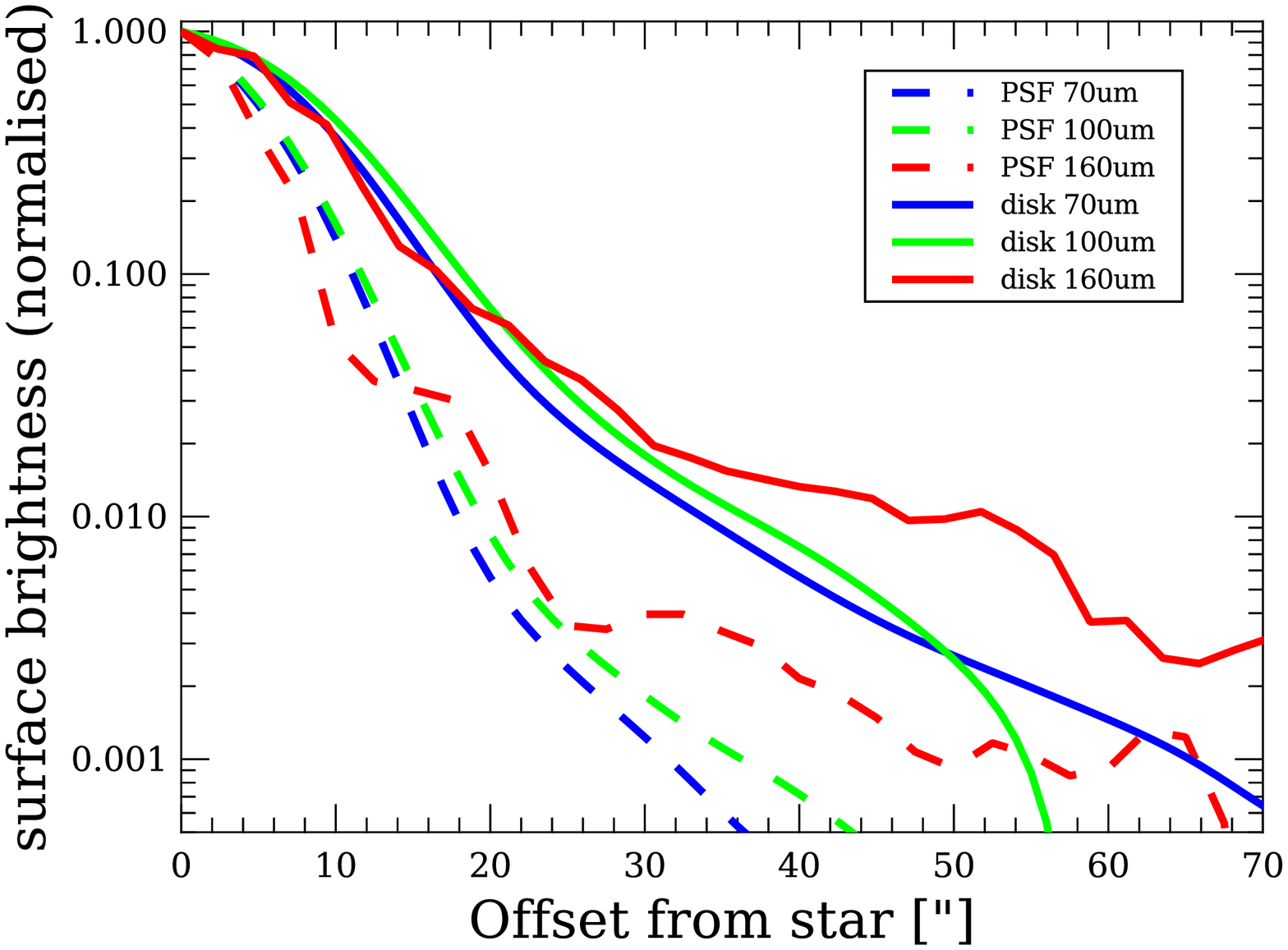}
 \caption{Normalised surface brightness profiles along the disk in NE direction in 
          the three PACS filters. The profiles were convolved with a Gaussian to 
	  match the spatial resolution of the 160\,$\mu$m image. The same convolved
	  profiles are shown for the PSF maps.  
	  }
 \label{Fig:profile_lambda}
\end{figure}

The comparison of the surface brightness profiles in the three PACS filters 
in Fig.~\ref{Fig:profile_lambda} shows the same brightness profile along the 
30.8\degr\ position angle in NE direction.  The 70 and 100\,$\mu$m profiles 
were convolved with a Gaussian to match the spatial resolution at 160\,$\mu$m.
The same convolution was applied to the 70 and 100\,$\mu$m PSF profiles.
The shape of these convolved PSF profiles defers significantly from that of the
160\,$\mu$m PSF profile. The
wiggles in the 160\,$\mu$m profile differ up to a factor of 3 from the convolved
70 and 100\,$\mu$m PSF profiles.
Within these uncertainties,  
there is no evidence of a wavelength dependent surface brightness. 
This indicates that the grains producing the emission at 70, 100, and 160\,$\mu$m 
are confined to the same locus in the disk.
At 70\,$\mu$m, the broadening of the profile with respect to the PSF indicates
that 90\% of the emission originates in a region within 11\arcsec\ or 200\,AU of the star.

\section{The far-infrared SED and grain size}

We integrated the surface brightness maps over a 60\arcsec\ radius circular
aperture.  Background subtraction was based on a rectangular region, selected
close enough to the object to be within the map region with the same coverage
as the center of the map.  For the background outlier rejection, the DAOphot
algorithm in the HIPE aperture photometry task was used.  The aperture 
photometry obtained provides a good
measure of the flux density of the integrated disk. The contribution of 
the stellar photosphere at these wavelengths is negligible.  The error is dominated
by the present uncertainties in the absolute flux calibration of both 
instruments.  The full disk flux densities are listed in 
Table~\ref{table:measuredQuantities}.

\begin{figure}
 \centering
 \includegraphics[width=9.3cm]{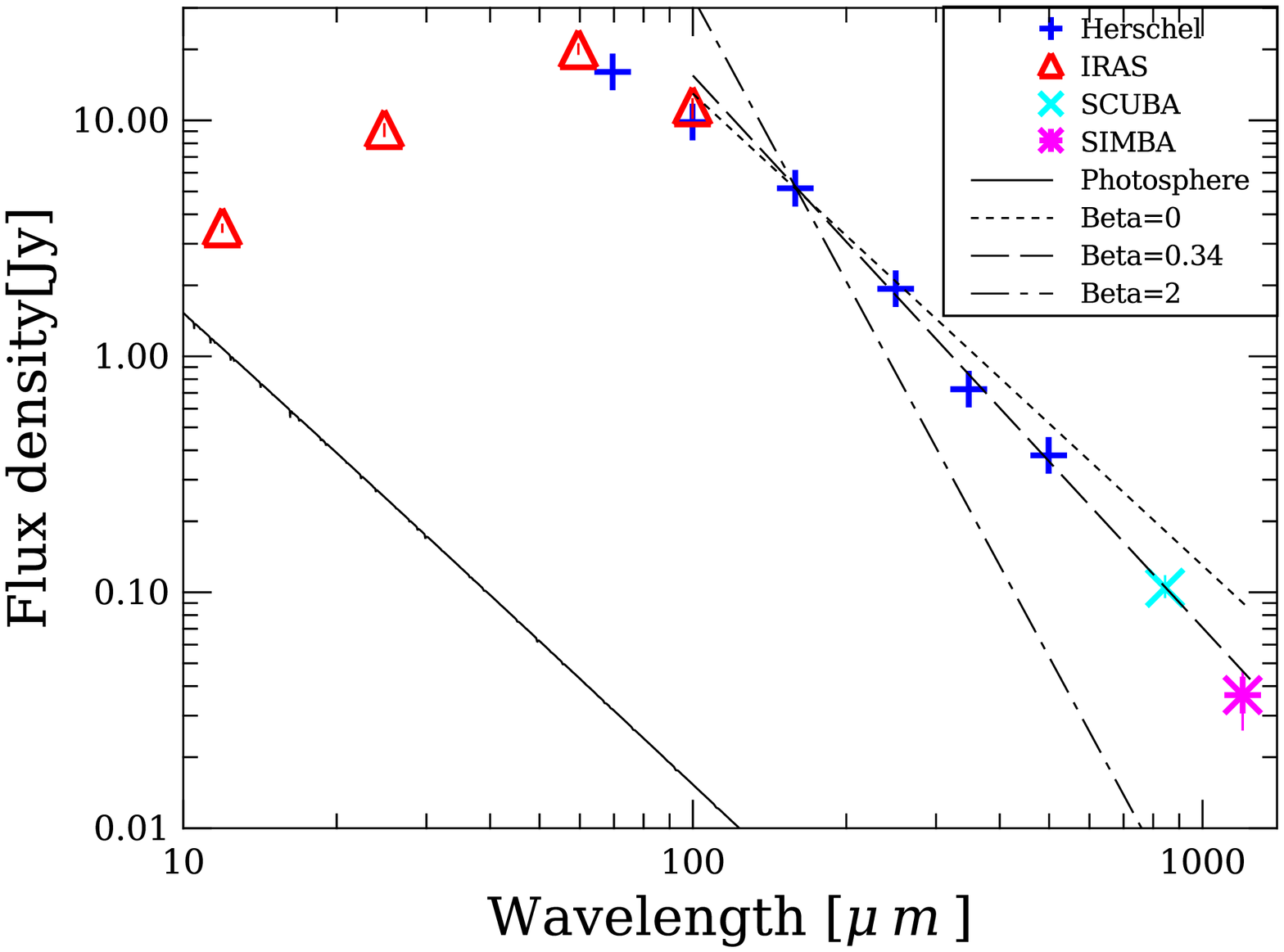}
 \caption{The infrared to mm SED of $\beta$\,Pic.  The PACS
          (70 -- 160\,$\mu$m), and the SPIRE 
	  (250 -- 500\,$\mu$m) fluxes were integrated over 
	  a 60\arcsec\ radius aperture.
	  IRAS flux densities are from the IRAS Point Source Catalog. 
	  The 850\,$\mu$m SCUBA datum \citep{holland:1998} and
	  the 1200\,$\mu$m SIMBA datum \citep{liseau:2003} are 
	  integrated over a 40\arcsec\ radius aperture.
          Overplotted is a Rayleigh-Jeans extrapolation of 
	  the 160\,$\mu$m flux density with a spectral index $\beta$=0 and $\beta$=2,  
	  and the best fit to the 160--1200\,$\mu$m data ($\beta$=0.34).
	  The stellar photosphere is a Kurucz model for $T_{\mathrm{eff}}$=9000\,K;log(g)=3.9 scaled to the 
	  2MASS photometry K$_s$=3.52.
	  }
 \label{Fig:betapicSED}
\end{figure}

Figure~\ref{Fig:betapicSED} shows the new PACS and SPIRE photometry, and selected 
infrared and (sub-)mm flux densities from the literature.  
Because the disk is optically thin at these wavelengths, the 
wavelength dependence of the emission directly probes the 
dust grains, and, in particular, their size distribution. 
We overplot two modified Rayleigh-Jeans laws ($F_\nu$ $\propto$ $\nu^{(2+\beta)}$),
normalized to the 160\,$\mu$m datum.  The spectral index $\beta$ indicates the 
mean dust opacity $\kappa \propto \nu^{\beta}$. An index
$\beta=0$ corresponds to a black body with a $\kappa$ independent of
wavelength $\lambda$, 
indicating grains that are much larger than $\lambda/2\pi$. 
Interstellar grains, which have a size distribution 
f(a)\,$\propto$\,a$^{-q}$ with $q=3.5$ and an upper size limit of
a$_{\rm max} \sim 0.3\,\mu$m,
are characterized by $\beta=1.8\,\pm\,0.2$ \citep{draine:2006}.
In protoplanetary disks, $\beta$-values from 1.5 down to 0 are found,
depending on the disk geometry \citep{acke:2004}.
An error-weighted least squares fit of a Rayleigh-Jeans law to 
the $\beta$\,Pic photometry at wavelengths beyond 160\,$\mu$m
yields $\beta\,=\,0.34\,\pm\,0.07$.   \citet{nilsson:2009} 
obtained $\beta\,=\,0.67$ from a $\beta$-corrected black-body fit to the full 
disk SED, including mid-infrared photometry.  The difference between both results
should not be over-interpreted since both approaches are sensitive in different ways
to simplifying assumptions about the temperature and size distribution within the disk.  In any 
case, both results consistently show a value below 0.7. 
\citet{ricci:2010} demonstrate that such a low value cannot be explained with a 
$q\,=\,3.5$ power law.
This is a surprise insofar as the latter value is the expected result for a 
population of bodies in a standard steady-state collisional cascade \citep{dohnanyi:1969}.

The grain size distribution in $\beta$\,Pic must be flatter than the $q\,=\,3.5$ power 
law, meaning that the fraction of small particles must be lower.  
Radiation  pressure can push the smallest grains (with 
$F_{\rm rad}/F_{\rm grav}>0.5$) onto hyperbolic orbits, hence reduce the
time these particles spend in the inner part of the disk, which can decrease
their volume density by two orders of magnitude \citep{krivov:2000}.
The disk cannot be fully cleared of small particles, since it has been seen 
in scattered light out to 1800\,AU.  The scattering grains are
probably the (sub-)$\mu$m grains that are blown out of the inner
disk, where the collisions take place.
However, this effect only reduces the densities for grains of size below
a few micrometers, and even fully removing these grains would not change 
$\beta$ to the observed value.

The small value of $\beta$ can be interpreted in a number of ways.  
The grain size distribution can exhibit a wavy pattern, caused by  
the absence of impactors small enough to be efficiently blown
out of the disk by radiation pressure.  This causes an over-abundance
of the grains that are just bound, which means there are more impactors
for the next larger size population. The reduction of this population 
causes an over-abundance of a following size population and so on
\citep{krivov:2006}.
The wavy size distribution can lead to small values of $\beta$ when measured
in the FIR \citep{thebault:2007}.  If the wavy structure were as strong
as found in this paper for normal and weak material properties, it would be
consistent with the small $\beta$ value we have measured.  
However, the strength and phase of the wavy pattern in the size distribution 
depend on both the grain structure and the eccentricity of the dust orbits in the disk. 

Alternative explanations of the small value of $\beta$ cannot be excluded.  
There are indications that the grains produced 
in the deep impact experiment followed a flatter power law with $q \approx 3.1$ 
\citep{jorda:2007}. 
Laboratory experiments illustrate that fragments produced in collisions of porous
aggregates can follow much flatter slopes 
\citep[$q=1.2$,][]{guttler:2009}, demonstrating that the porosity of the colliding 
grains should not be disregarded.

Additional dynamical models should be developed to quantify the possible
contribution of these effects to the small $\beta$ observed in
$\beta$\,Pic.

\section{Conclusions}

We have presented images of the $\beta$\,Pic debris disk 
in six photometric bands between 70 and 500\,$\mu$m using the
PACS and SPIRE instruments.  
We resolve the disk at 70, 100, 160, and 250\,$\mu$m.
The images at 70--160\,$\mu$m show no evidence of asymmetries in the far-infrared
surface brightness along the disk of $\beta$\,Pic.
The observed profiles are compatible 
with 90\% of the emission originating in a region within a radius of 200\,AU from the star. 
The disk-integrated photometry in the six \emph{Herschel} filters provides a far
infrared SED with small spectral index $\beta\,\approx\,0.34$, which is indicative of
a grain size distribution that is inconsistent with a local collisional equilibrium.
The size distribution is modified by either non-equilibrium effects, or 
exhibits a wavy pattern, caused by the under-abundance of
impactors that are small enough to be removed by radiation pressure.

\begin{acknowledgements}
PACS has been developed by a consortium of institutes led by MPE (Germany) 
and including UVIE (Austria); KU Leuven, CSL, IMEC (Belgium); 
CEA, LAM (France); MPIA (Germany); INAF-IFSI/OAA/OAP/OAT, LENS, SISSA (Italy); 
IAC (Spain). This development has been supported by the
funding agencies BMVIT (Austria), ESA-PRODEX (Belgium), CEA/CNES (France), DLR (Germany),
ASI/INAF (Italy), and CICYT/MCYT (Spain).
SPIRE has been developed by a consortium of institutes led by
Cardiff Univ. (UK) and including Univ. Lethbridge (Canada);
NAOC (China); CEA, LAM (France); IFSI, Univ. Padua (Italy);
IAC (Spain); Stockholm Observatory (Sweden); Imperial College
London, RAL, UCL-MSSL, UKATC, Univ. Sussex (UK); Caltech, JPL,
NHSC, Univ. Colorado (USA). This development has been supported
by national funding agencies: CSA (Canada); NAOC (China); CEA,
CNES, CNRS (France); ASI (Italy); MCINN (Spain); SNSB (Sweden);
STFC (UK); and NASA (USA). 
BV acknowledges the Belgian Federal 
Science Policy Office via the ESA-PRODEX office.
The authors thank the referee for several helpful comments.
\end{acknowledgements}

\bibliographystyle{aa}

\Online

\appendix
\section{Data reduction}
\label{section:datareduction}

The PACS data were processed in the \emph{Herschel} interactive analysis
environment HIPE (v3.0), applying the standard pipeline steps. 
The flux conversion was done using version 5 of the response calibration.
Signal glitches due to cosmic ray impacts were masked out in two steps.
First the PACS photMMTDeglitching task in HIPE was applied on the detector
timeline. Then a first coarse map was projected, which is then used
as a reference for the second level deglitching HIPE task IIndLevelDeglitch.
In the detector time series we masked the region around the source
prior to applying a high-pass filter to remove the low frequency drifts.
The scale of the high pass filter was taken to be half the length of 
an individual scan leg on the sky, i.e. 3.7\arcmin.  
The detector time series signals were then summed up into a map using 
the PACS photProject task.  The pixel scale for the 70 and 100\,$\mu$m
maps was set to 1\arcsec, while the scale for the 160\,$\mu$m map was 2\arcsec.
For the deep map in the 70 and 160\,$\mu$m filter we combined
the two detector time series and projected these together.

The SPIRE data were also reduced using HIPE and maps
were obtained via the default naiveMapper task. The SPIRE observation
consists of several repetitions of a map observation of the same
area.  As a result it was possible to project the data with a pixel size of 
4, 6, and 9\arcsec\ while still maintaining complete sampling across the source. 

\end{document}